\documentclass[%
aps,
prd,
preprint,
nofootinbib,
superscriptaddress,
amsmath,
amssymb]{revtex4-1}

\usepackage{indentfirst}  
\usepackage{bm}    
\usepackage{graphicx}  
\usepackage{latexsym}
\usepackage{graphicx}
\usepackage{psfrag}
\usepackage{epsfig}
\usepackage{amsmath}
\usepackage{tocloft}
\usepackage{array}
\usepackage{setspace}
\usepackage{amssymb}
\usepackage{float}
\usepackage{multirow}
\usepackage{slashbox}
\usepackage{color}

\begin{document}

\title{Quasi-B-mode generated by high-frequency gravitational waves and corresponding perturbative photon fluxes}   



\author{Fangyu Li}%
\email[]{cqufangyuli@hotmail.com}
\affiliation{Institute of Gravitational Physics, Department of Physics, Chongqing University, Chongqing 400044, P.R. China}

\author{Hao Wen}
\affiliation{Institute of Gravitational Physics, Department of Physics, Chongqing University, Chongqing 400044, P.R. China}

\author{Zhenyun Fang}%
\affiliation{Institute of Gravitational Physics, Department of Physics, Chongqing University, Chongqing 400044, P.R. China}

\author{Lianfu Wei}%
\affiliation{Quantum Optoelectronics Laboratory, Southwest Jiaotong University, Chengdu 610031, China}
\author{Yiwen Wang}%
\affiliation{Quantum Optoelectronics Laboratory, Southwest Jiaotong University, Chengdu 610031, China}
\author{Miao Zhang}%
\affiliation{Quantum Optoelectronics Laboratory, Southwest Jiaotong University, Chengdu 610031, China}


\date{\today}

\begin{abstract}
Interaction of very low-frequency primordial (relic) gravitational waves (GWs)to cosmic microwave background (CMB)can generate B-mode polarization. Here, for the first time we point out that the electromagnetic (EM)response to high-frequency GWs (HFGWs)would produce quasi-B-mode distribution of the perturbative photon fluxes. We study the duality and high complementarity between such two B-modes, and it is shown that such two effects are from the same physical origin: the tensor perturbation of the GWs and not the density perturbation. Based on this quasi-B-mode in HFGWs and related numerical calculation, it is shown that the distinguishing and observing of HFGWs from the braneworld would be quite possible due to their large amplitude, higher frequency and very different physical behaviors between the perturbative photon fluxes and background photons, and the measurement of relic HFGWs may also be possible though face to enormous challenge.\\


\end{abstract}

\maketitle

\section{Introduction}
\label{Introduction}

\indent On  11th Feb 2016 and June 2016, LIGO reported two very important evidences\cite{PhysRevLett.116.061102,secondLIGOGW} of GW detection. One of them is the GW having amplitude of $h\sim 10^{-21}$ and frequency of $\sim35$ to $350Hz$. Another one is the GW with amplitude of $h\sim10^{-22}$ and frequency of $\sim35$ to $450Hz$, and they are all produced by black hole mergers, which come from distance of 1.3 and 1.4 billion light-years from the Earth, respectively.
Obviously, such results are very big encouraging to GW project, and they should push forward research of GW projects, including observation and detection for GWs in different frequency bands, different kinds of GWs, and in different ways. Thus, they should be highly complementary each other.

Before this, in 2014, observation of the B-mode polarization caused by primordial(relic) gravitational
waves(GWs) in the cosmic microwave background(CMB) has been reported\cite{PhysRevLett.112.241101}. If this B-mode polarization can be
completely confirmed by experimental observation, it must also be a great encouragement for detection of GWs in the very-low
frequency band, and will provide a key evidence for the inflationary model.\\
\indent On the other hand, influence of cosmic dusts might swamp the signal of the B-mode polarization\cite{1475-7516-2014-10-035}.
In addition, if strength of these primordial GWs can reach up to the value reported by the B-mode experiment, then the temperature
perturbation induced by the primordial GWs should also be observed, but the Planck satellite did not observe such temperature
perturbation. Therefore,
further analysis to the B-mode polarization results by data of Planck satellite and other observation ways,
will provide critical judgement for the  B-mode polarization. However, no matter what the current result is, it
should not impact the scheme of observation for B-mode effect caused by the relic GWs, but should strongly attract further attentions
of scientific communities on this important phenomenon from the tensor perturbation, and in the future, it would be promising that
the research works of B-mode polarization will bring us crucial constraints on the inflationary models.\\
\indent It should be pointed out that almost all mainstream early universe models and inflation theories predicted primordial (relic) GWs, which have a very broad frequency band distribution. During the very early universe and the inflation epoch of the universe, since extreme small spacetime scale and huge high energy density (they are close to the Planck scale), the quantum effect would play important role and might provide important contribution to generation of the relic gravitons. Then Heisenberg principle would govern the creation and the annihilation of the particles. In this case severe quantum fluctuation would have pumped huge energy into the production of gravitons. In this period, the gravitons having huge energy correspond to extreme-high frequency.\\
\indent However,the rapid expansion of the universe would have stretched the graviton wavelengths from microscopic to macroscopic length, and present values of these graviton wavelengths would be expected to be from $\sim1cm$ to the cosmological scale. In other words, the frequency spectrum of the relic GWs would be from $\sim10^{-17}Hz$ to $\sim10^{10}Hz$, roughly. Nevertheless, the spectrum densities and dimensionless amplitudes expected by different universe models and scenarios are different due to the different cosmological parameters. Moreover, string theory\cite{pr373}, loop quantum gravity\cite{copeland2008} and some classical and semi-classical scenarios\cite{prd044017,Kogun_CQG21_2004} also expected the HFGWs, and some of them have interesting and significant strength and properties. Frequency band of the relic GWs predicted by the ordinary inflationary models
\cite{0504018,084022},
the quintessential inflationary model\cite{prd123511,cqg045004, Giovannini2014} and the pre-big-bang model\cite{pr373,sciam290} have been extended to very high frequency range
($\sim10^8~to~10^{10}~Hz$). Moreover, high-frequency  GWs(HFGWs) expected by the braneworld senarios\cite{cqgF33} and
interaction
of astrophysical plasma with intense electromagnetic(EM) radiation from high-energy astrophysical process\cite{prd044017} have been extended to $\sim10^9~Hz~
to~10^{12}~Hz$ or higher frequency, and corresponding dimensionless amplitudes of these HFGWs might reach up to $h\sim
10^{-22}~to~10^{-27}$(see Table I)\cite{prd044017,cqgF33}. Besides,  even high-energy physics
experiments\cite{Chen1994,PhysRevD.78.094002}[e.g., see our previous work: Large Hadron Collider(LHC)] also predicted extremely-high frequency GWs(high-energy gravitons)\cite{PhysRevD.78.094002},
and their frequencies might reach up to $10^{19}$ to $10^{23}$ Hz, but the dimensionless amplitude may be only $\sim10^{-39}$ to
$10^{-41}$. Obviously the frequencies  of these HFGWs are far beyond the detection  or observation range
of the intermediate-frequency GWs(e.g., LIGO, GEO600, Virgo, TAMA\cite{Abbott_PRD062002_2009,12,virgo,TAMA,GEO600}, $\nu\sim1$ to $10^4$Hz), the low-frequency GWs detection(e.g., LISA, 
BBO, DECIGO....\cite{LISA,BBO,DECIGO}, $\nu\sim10^{-7}$ to $1$Hz),and very low-frequency GWs($\nu\sim10^{-16}$ to $10^{-17}$Hz, e.g., B-mode experiment in the CMB). Thus, detection and
observation of these HFGWs need new principle and scheme. Once the HFGWs can be detectable and observable, then which will open
a new information window into the cosmology and the high-energy astrophysical process, and would be highly complementary for
the observation of the GWs in the intermediate-, the low-frequency and the very low-frequency bands.\\


\begin{table}[!htbp]
	\scriptsize
	\caption{\label{tab:Table I}%
		Some of possible HFGWs and related properties.}
	\begin{tabular}{cccccc}
		\hline
		\hline
		Possible&Ordinary &Quintessential&Pre-big-bang
		&Brane&Interaction of \\
		
		HFGWs&inlationary\cite{0504018,084022}& inflationary\cite{prd123511,cqg045004,Giovannini2014}&\cite{pr373,sciam290}
		&Oscillation\cite{cqgF33}&astrophysical plasma \\
		
		&&&&& with intense \\
		&&&&&EM radiation\cite{prd044017}\\
		\hline
		\\
		Frequency bands&$\sim10^8-10^{10}Hz$&$\sim10^9-10^{10}Hz$&$\sim10^9-10^{10}Hz$&$\sim10^8-10^{14}Hz$&$\sim10^9-10^{12}Hz$\\
		\\
		Dimensionless &$\sim10^{-30}(upper~limit)$&$\sim10^{-30}-10^{-31}$&$\sim10^{-29}-10^{-31}$&
		$\sim10^{-22}-10^{-25}$&$\sim10^{-25}-10^{-27}$\\
		amplitudes&$-10^{-34}$ or less&~&~&~&~\\
		&Stochastic&Stochastic&Stochastic&Discrete&Continuous\\
		Properties&background&background&background&spectrum&spectrum\\
		\hline
	\end{tabular}
\end{table}


\indent It should be pointed out that the tensor perturbation of GWs is a very common property, which can be not only expressed
as B-mode polarization\cite{PhysRevLett.112.241101,science989} in the CMB for very low-frequency relic GWs, but also quasi-B-mode distribution of perturbative
photon fluxes in electromagnetic response for HFGWs.
However, the duality and similarity between the B-mode of the CMB experiment for the very low frequency GWs and the quasi-B-mode
of electromagnetic(EM) response for the HFGWs, almost have never been studied in the past.
In fact, these effects are all from  the
same physical origin: tensor perturbation of the GWs and not the density perturbation, and they would be highly complementary
, not only in the observable frequency bands, but also in the displaying ways. \\
\indent In this paper we shall study the similarity and duality between the B-mode polarization in the CMB for very
low frequency primordial GWs and the quasi-B-mode distribution of the perturbative photon fluxes(i.e., signal photon
fluxes) in the EM response for HFGWs. It is shown that such two B-modes have a fascinating duality and strong complementarity,
and distinguishing and observing of the HFGWs expected by the braneworld would be quite possible due to their large amplitude,
higher frequency and very different physical behaviors between the perturbative photon fluxes and the background photon fluxes. The measurement of
relic HFGWs may also be possible though it face to enormous challenge.\\
\indent The plan of this paper is follows. In Sec.II we study the strength and angular distribution of the perturbative photon fluxes generated by
the HFGWs expected by some typical cosmological models and high-energy astrophysical process, and discuss the duality and similarity
between such two B-modes, especially their complementarity due to the same physical reason: the tensor perturbation. In Sec.III
we consider displaying conditions for the HFGWs, including the quasi-B-mode experiment in the EM response for the HFGWs.
In Sec.IV we discuss wave impedance and wave impedance matching to the perturbative photon fluxes and the background photon fluxes. Our brief conclusion
is summarized in Sec.V.

\section{Quasi-B-mode in electromagnetic response to the high-frequency GWs}
\indent It is well known that, ``monochromatic components'' of the GWs propagating along the z-direction can often be written
as \cite{Maggiore}
\begin{eqnarray}\label{eq1}
	&~&h_{\mu\nu} =\left( {{\begin{array}{*{20}c}
				0 &  0   & 0 &0\\
				0 & A_{\oplus}&A_{\otimes}  &0\\
				0 & A_{\otimes}  &  -A_{\oplus}& 0\\
				0 &  0   & 0&0 \nonumber\\
			\end{array} }} \right)exp[i(k_gz-\omega_gt)]\nonumber\\
		\end{eqnarray}
		\indent For the relic GWs, $ A_{\oplus}=A(k_g)/a(t)$, $ A_{\otimes}=A(k_g)/a(t)$, \cite{0504018,084022} are the stochastic values of the
		amplitudes of the relic GWs in the laboratory frame of reference, $\oplus$ and $\otimes$ represent the  $\oplus$-type and
		$\otimes$-type polarizations, and $k_g$, $\omega_g$ and $a(t)$ are wave vector, angular frequency and the cosmology scale factor
		in  the laboratory frame of reference, respectively. For the non-stochastic coherent GWs,  $A_{\oplus}$ and $A_{\otimes}$ are
		constants.\\
		According to Eq.(\ref{eq1}) and electrodynamic equation in curved spacetime, the perturbative EM fields produced by the
		direct interaction of the incoming GW, Eq.(\ref{eq1}), with a static magnetic field $\hat{B}^{(0)}$, can be
		give by\cite{nuovo129,FYLi_EPJC_2008}(we use MKS units)
		\begin{eqnarray}
			\label{eq2}
			{ { \tilde {E}_x^{(1)}}}&=& -{\frac{ \textit{i}}{2}}  A_{\oplus} { { \hat{B}}}^{(0)}k_g c \Delta \textit{l} \exp [i(k_g z-\omega_g t)]  ,\nonumber\\
			{ { \tilde {B}_y^{(1)}}}&=& -{\frac{ \textit{i}}{2}}  A_{\oplus} { { \hat{B}}}^{(0)}k_g  \Delta \textit{l} \exp [i(k_g z-\omega_g t)] ,\nonumber\\
			{ { \tilde {E}_y^{(1)}}}&=& -{\frac{ i}{ 2}}  A_{\otimes} { { \hat{B}}}^{(0)}k_g c \Delta l \exp [i(k_g z-\omega_g t)],\nonumber\\
			{ { \tilde {B}_x^{(1)}}}&=& ~~\frac{i}{2}A_{\otimes} { { \hat{B}}}^{(0)}k_g \Delta l \exp [i(k_g z-\omega_g t)],
		\end{eqnarray}
		where $\Delta\textit{l}$ is the interaction dimension between the HFGW and the static magnetic field $\hat{B}^{(0)}$, which is
		perpendicular to the propagating direction of the HFGW, ``$\wedge$'' stands for the static background magnetic field,
		``$\sim$'' represents time-
		dependent perturbative EM fields, and the superscript(0) and (1) denote the background and the first-order perturbative EM fields,
		respectively. Here the perturbative EM fields propagating along the negative z direction(i.e., the opposite propagation direction
		of the HFGW) are neglected, because they are much weaker or absent\cite{nuovo129, FYLi_EPJC_2008,prd2915}. We shall show that using
		EM synchro-resonance
		($\omega_e=\omega_g$) system of coupling between the static magnetic field $\hat{B}^{(0)}$ and a Gaussian type-photon flux(the Gaussian
		beam),  the ``quasi-B-mode'' of strength distribution of the perturbative photon flux and the B-mode polarization in the
		CMB have interesting duality and they would be highly complementary.\\
		\indent    According to the quantum electronics, form of the Gaussian-type photon fluxes[the Gaussian beam] is
		actually expressed by wave beam solution from the Helmholtz equation, and the most basic and general
		form of the Gaussian beams
		is the elliptic mode of fundamental frequency\cite{Yariv}, i.e,
		\begin{eqnarray}
			\label{eq3}
			\psi &=&\frac{\psi _0\cdot\exp [-(\frac{x^2}{W_x^2}+\frac{y^2}{W_y^2})]}{\sqrt {[1+(z-z_x)^2/f_x^2]^{\frac{1}{2}}\cdot[1+(z-z_y)^2/f_y^2]^{\frac{1}{2}}}}\nonumber\\
			&\cdot&\exp
			\{i[(k_e z-\omega _e t)-\frac{1}{2}[\tan^{-1}(\frac{z-z_x}{f_x})+\tan^{-1}(\frac{z-z_y}{f_y})]
			+\frac{k_e}{2}(\frac{x^2}{R_x}+\frac{y^2}{R_y})+\delta ]\},\nonumber\\
		\end{eqnarray}
		where $\psi_0$ is the amplitude of the Gaussian beam, $R_x=z+f_x^2/(z-z_x)$ and $R_y=z+f_y^2/(z-z_y)$ are the curvature radii of the wave fronts at the xz-plane and at the
		yz-plane of the Gaussian beam,  $f_x=\pi W_{0 x}^2/\lambda_e$, $f_y=\pi W_{0 y}^2/\lambda_e$, $W_x=W_{0 x}[1+(z-z_x)^2/f_x^2]^{\frac{1}{2}}$,
		$W_y=W_{0 y}[1+(z-z_y)^2/f_y^2]^{\frac{1}{2}}$,  $W_{0 x}$ and $W_{0 y}$  are the minimum spot radii of the Gaussian beam at
		the xz-plane and at the yz-plane, respectively. Here, we shall study case of $R_x=R_y=R$, $z_x=z_y=z$, $W_x=W_y=W$ and
		$f_x=f_y=f$, i.e., then the elliptic Gaussian beam, Eq.(\ref{eq3}), will be reduced to the circular Gaussian beam\cite{Yariv}.\\
		\indent By using the condition of non-divergence  $\nabla\cdot\bf{\tilde{E}}^{(0)}=0$ in free space and
		$ \rm{\bf\tilde{B}^{(0)}}=-i/\omega_e \bigtriangledown\times \rm{\bf\tilde{E}^{(0)}}$, we find
		a group of special wave beam solution of the Gaussian beam as follows:
		\begin{eqnarray}
			\label{eq4}
			{\tilde{E}}_x^{(0)}=\psi_{ex}&=&\psi,
			{\tilde{E}}_y^{(0)}=\psi_{ey}=0,\nonumber\\
			{\tilde{E}}_z^{(0)}=\psi_{ez}&=&2x\int(\frac{1}{W^2}-i\frac{k_e}{2R})\psi dz
			=2xF(\mathbf{x},k_e,W),\nonumber\\
			F(\mathbf{x},k_e,W)&=&\int(\frac{1}{W^2}-i\frac{k_e}{2R})\psi dz,
		\end{eqnarray}
		\begin{eqnarray}
			\label{eq5}
			{\tilde{B}}_x^{(0)}&=&\psi_{bx}=-\frac{i}{\omega_e}\frac{\partial\psi_{ez}}{\partial y},
			{\tilde{B}}_y^{(0)}=\psi_{by}=-\frac{i}{\omega_e}(\frac{\partial\psi}{\partial z}-\frac{\partial\psi_{ez}}{\partial x}),
			{\tilde{B}}_z^{(0)}=\psi_{bz}=\frac{i}{\omega_e}\frac{\partial\psi}{\partial y}.\nonumber\\
		\end{eqnarray}

		Here ${\tilde{B}}_z^{(0)}=\psi_{bz}$ is a crucial parameter since the strength and physical behaviour of transverse
		perturbative photon flux(the transverse signal photon fluxes) mainly depend on ${\tilde{B}}_z^{(0)}$[see below
		and Eqs.(\ref{eq11}) to (\ref{eq13})]. Using Eqs.(\ref{eq3}) and (\ref{eq5}), we have\\
		\begin{eqnarray}
			\label{eq6}
			&~&{\tilde{B}}_z^{(0)} = -[\frac{\psi_0k_e r sin\phi}{\omega_e[1+(z/f)^2]^{\frac{1}{2}}(z+f^2/z)}+\frac{i2\psi_0rsin\phi}
			{\omega_eW_0^2[1+(z/f)^2]^{\frac{3}{2}}}]\nonumber\\
			&\cdot&\exp(-\frac{r^2}{W^2})\exp{\{i[(k_e z-\omega _e t)-\tan ^{-1}(\frac{z}{f})+\frac{k_e r^2}{2R}+\delta ]\}}.
		\end{eqnarray}\mbox{}
		From Eqs.(\ref{eq3}) to (\ref{eq6}), we obtain the strength of the transverse background photon fluxes in cylindrical polar
		coordinates as follows:
		\begin{eqnarray}
			\label{eq7}
			&~&n_{\phi}^{(0)}=-n_x^{(0)}\sin\phi+n_y^{(0)}\cos\phi
			=-\frac{c}{\hbar\omega_e}\langle \mathop{T^{01}}\limits^{(0)} \rangle\sin
			\phi+ \frac{c}{\hbar\omega_e}\langle \mathop{T^{02}}\limits^{(0)} \rangle\cos\phi\nonumber\\
			&=&\frac{1}{2\mu_0\hbar\omega_e}Re\langle\psi_{ez}^*\psi_{by}\rangle\sin\phi
			+\frac{1}{2\mu_0\hbar\omega_e}Re\langle\psi^*\psi_{bz}\rangle\cos\phi\nonumber\\
			&+&\frac{1}{2\mu_0\hbar\omega_e}Re\langle\psi_{ez}^*\psi_{bx}\rangle\cos\phi
			=f_{\phi}^{(0)}\exp(-\frac{2r^2}{W^2})\sin2\phi,
		\end{eqnarray}
		where  $\mathop{T^{01}}\limits^{(0)}$ and $\mathop{T^{02}}\limits^{(0)}$ are 01- and 02-components of the energy-momentum
		tensor for the background EM wave(the Gaussian beam), and
		\begin{eqnarray}
			\label{eq8}
			n_x^{(0)}&=&\frac{1}{2\mu_0\hbar\omega_e}Re\langle\psi_{ez}^*\psi_{by}\rangle
			=\frac{1}{2\mu_0\hbar\omega_e}Re\langle{\tilde{E}}_z^{(0)*}{\tilde{B}}_{y}^{(0)}\rangle\\
			\label{eq9}
			n_y^{(0)}&=&\frac{1}{2\mu_0\hbar\omega_e}Re[\langle\psi^*\psi_{bz}+\langle\psi_{ez}^*\psi_{bx}\rangle]
			=\frac{1}{2\mu_0\hbar\omega_e}Re\langle[{\tilde{E}}_x^{(0)*}{\tilde{B}}_{z}^{(0)}+{\tilde{E}}_z^{(0)*}{\tilde{B}}_{x}^{(0)}
			]\rangle \nonumber\\
		\end{eqnarray}
		are the transverse background photon fluxes in the x-direction and in the y-direction, respectively, where $*$ denotes
		complex conjugate, and the angular brackets represent the average over time, and(also see Ref.\cite{Yariv})
		\begin{eqnarray}
			\label{eq10}
			n_x^{(0)}|_{x=0}=n_y^{(0)}|_{y=0}=0.
		\end{eqnarray}
		From Eqs.(\ref{eq3}) to (\ref{eq10}), we obtain the strength distribution of $n_{\phi}^{(0)}$ as follows(see Fig.1)
		\begin{figure}[!htbp]
			\centerline{\includegraphics[scale=0.8]{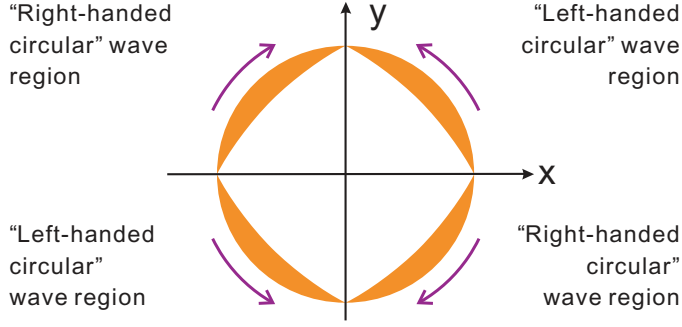}}\label{fig1}
			\caption{\footnotesize{The strength distribution of background photon flux $n_{\phi}^{(0)}$, Eq.(\ref{eq7}), in the cylindrical polar
					coordinates.}}
		\end{figure}
		
		\indent In the same way, under the resonance condition($\omega_e=\omega_g$), from Eqs.(\ref{eq2}),(\ref{eq3}),(\ref{eq6}), the
		transverse perturbative photon flux(the signal photon flux) can be given by:
		\begin{eqnarray}
			\label{eq11}
			&~&n_{\phi}^{(1)}=-n_{x}^{(1)}\sin\phi+n_{y}^{(1)}\cos\phi
			=-\frac{c}{\hbar\omega_e}\langle \mathop{T^{01}}\limits^{(1)} \rangle_{\omega_e=\omega_g}\sin
			\phi+ \frac{c}{\hbar\omega_e}\langle \mathop{T^{02}}\limits^{(1)} \rangle_{\omega_e=\omega_g}\cos\phi\nonumber\\
			&=&-\frac{1}{2\mu_0\hbar\omega_e}Re\langle{\tilde{E}}_y^{(1)*}{\tilde{B}}_{z}^{(0)}\rangle_{\omega_e=\omega_g}\sin\phi
			+\frac{1}{2\mu_0\hbar\omega_e}Re\langle{\tilde{E}}_z^{(0)*}{\tilde{B}}_{x}^{(1)}\rangle_{\omega_e=\omega_g}\cos\phi\nonumber\\
			&+&\frac{1}{2\mu_0\hbar\omega_e}Re\langle{\tilde{E}}_x^{(1)*}{\tilde{B}}_{z}^{(0)}\rangle_{\omega_e=\omega_g}\cos\phi
			=n_{\phi-\otimes}^{(1)}+n_{\phi-\otimes}^{(1)'}+n^{(1)}_{\phi-\oplus},
		\end{eqnarray}
		where $\langle \mathop{T^{01}}\limits^{(1)} \rangle_{\omega_e=\omega_g}$ and
		$\langle \mathop{T^{02}}\limits^{(1)} \rangle_{\omega_e=\omega_g}$ are average values of 01- and 02-components of energy-momentum
		tensor for first-order perturbation EM fields with respect to time, and
		
		\begin{eqnarray}
			\label{eq12}
			n_{\phi-\otimes}^{(1)}&=&\frac{1}{\mu_0\hbar\omega_e}\{
			\frac{A_{\otimes}\hat{B}^{(0)}\psi_0k_g\Delta \textit{l}r}{2[1+(z/f)^2]^{\frac{1}{2}}(z+f^2/z)}
			\sin[\frac{k_e r^2}{2R} -\tan ^{-1}(\frac{z}{f})+\delta]+\frac{A_{\otimes}\hat{B}^{(0)}
				\psi_0\Delta\textit{l}r}{W_0^2[1+(z/f)^2]^{\frac{3}{2}}}\nonumber\\
			&\cdot&\cos[\frac{k_e r^2}{2R} -\tan ^{-1}(\frac{z}{f})+\delta]\}exp(-\frac{r^2}{W^2})\sin^2\phi,
		\end{eqnarray}
		\begin{eqnarray}
			\label{eq13}
			n_{\phi-\otimes}^{(1)'}=\frac{1}{\mu_0\hbar\omega_e}\{\frac{1}{2}A_{\otimes}\hat{B}^{(0)}k_g\Delta \textit{l} Re \langle F^*(\textbf{x},k_g,W)
			\cdot exp[i(k_gz-\omega_gt+\pi/2)] \rangle\}_{\omega_e=\omega_g}\cos^2\phi,~~~
		\end{eqnarray}
		\begin{eqnarray}
			\label{eq14}
			n_{\phi-\oplus}^{(1)}&=&\frac{1}{\mu_0\hbar\omega_e}\{
			\frac{A_{\oplus}\hat{B}^{(0)}\psi_0k_g\Delta \textit{l}r}{4[1+(z/f)^2]^{\frac{1}{2}}(z+f^2/z)}
			\sin[\frac{k_e r^2}{2R} -\tan ^{-1}(\frac{z}{f})+\delta]\nonumber+\frac{A_{\oplus}\hat{B}^{(0)}
				\psi_0\Delta\textit{l}r}{2W_0^2[1+(z/f)^2]^{\frac{3}{2}}}\nonumber\\
			&\cdot&\cos[\frac{k_e r^2}{2R} -\tan ^{-1}(\frac{z}{f})+\delta]\}exp(-\frac{r^2}{W^2})\sin2\phi,
		\end{eqnarray}
		where $n_{\phi-\otimes}^{(1)}$ and $ n_{\phi-\otimes}^{(1)'} $ are the perturbative photon fluxes generated by the $\otimes$-type polarization state
		of the HFGW, and $n_{\phi-\oplus}^{(1)}$ is the perturbative photon flux produced by the  $\oplus$-type polarization state of the HFGW.\\
		\indent It is very interesting to compare the polarization patterns(see Fig.2a) in the CMB caused by primordial
		density perturbation\cite{science989} , the polarization patterns(see Fig.2b)\cite{science989,PhysRevLett.113.021301} produced by the relic GWs(in the very low frequency
		band) and the strength distribution(see Fig.2c and 2d) of the perturbative photon fluxes(in the high-frequency band), Eqs.(\ref{eq12}) to (\ref{eq13}),
		caused by the HFGWs, respectively.

		\begin{figure*}[!htbp]
			\centerline{\includegraphics[scale=0.7]{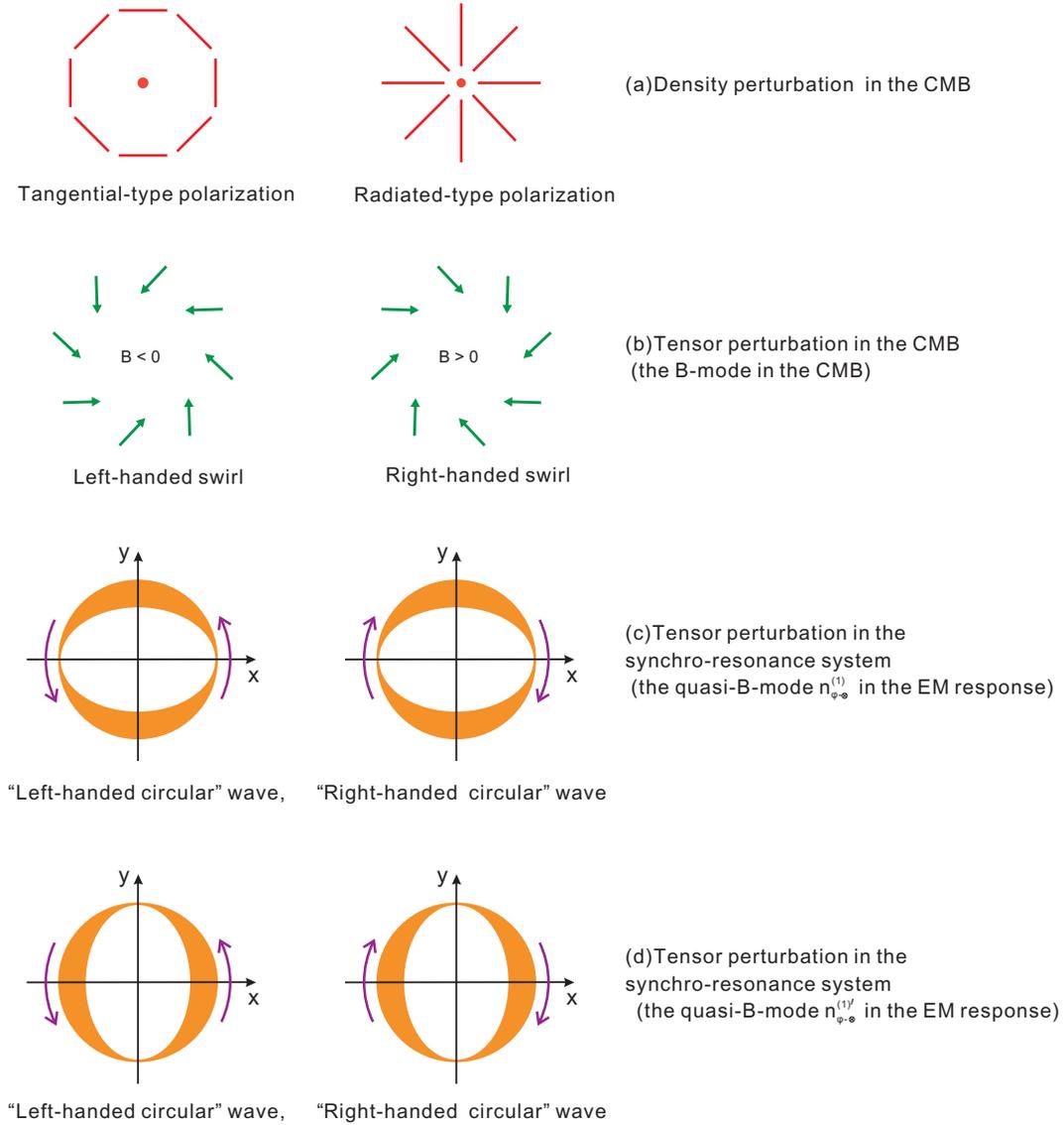}}\label{fig2}
			\caption{\footnotesize{The polarization patterns(Fig.2a) in the CMB caused by primordial density perturbation,
					the polarization patterns(Fig.2b) in the CMB produced by the relic GWs(tensor perturbation) in very low-frequency band, and the strength
					distribution(Fig.2c and 2d) of the perturbative photon fluxes in the EM response generated by the HFGWs(tensor perturbation)
					in the microwave frequency band.}}
		\end{figure*}

		\indent The density perturbation had no right-and-left handed orientation, thus their polarization are expressed as the
		tangential-type and radiated type patterns. Unlike the density perturbation(Fig.2a), the polarization patterns(Fig.2b) in the
		CMB produced by the relic GWs are the left-handed and right-handed swirls, and the EM response(Fig.2c and 2d)
		generated by the HFGWs in our synchro-resonance system are the ``left-handed circular wave'' and ``right-handed circular wave'', the latter both(Fig.2b,2c and 2d)
		are all from the tensor perturbation of the GWs. Here the ``left-handed circular'' or the ``right-handed circular'' property
		in the EM response depends on the phase factors in Eqs.(\ref{eq12}) and (\ref{eq13})(see Fig.3 and below).\\
		\indent By the way, the angular distributions of strength of the perturbative photon flux $n_{\phi-\oplus}^{(1)}$, Eq.(\ref{eq14}),
		and that of the background
		photon flux $n_{\phi}^{(0)}$, Eq.(\ref{eq7}), are the same(Fig.1), i.e.,
		they are not completely ``left-handed circular'' or completely ``right-handed circular''. In this case, $n_{\phi-\oplus}^{(1)}$
		will be swamped by $n_{\phi}^{(0)}$. Then, $n_{\phi-\oplus}^{(1)}$ has no observable effect, but $n_{\phi-\otimes}^{(1)}$ and
		$n_{\phi-\otimes}^{(1)'}$ would be observable(see below), and vice versa.
		Unlike $n_{\phi-\oplus}^{(1)}$, strength of $n_{\phi-\otimes}^{(1)}$, $n_{\phi-\otimes}^{(1)'}$ and $n_{\phi}^{(0)}$ have
		very different physical behaviours, such as different angular distribution and other properties. Eq.(\ref{eq12}) shows
		that $n_{\phi-\otimes}^{(1)}$ has maximum at $\phi=\pi/2$ and  $\phi=3\pi/2$(Fig.2c), and $n_{\phi-\otimes}^{(1)'}$ has maximum
		at $\phi=0$ and $\pi$(Fig.2d). This means that the peak value position of the signal photon fluxes are just the zero value
		areas($\phi=0$, $\pi/2$, $\pi$, $3\pi/2$) of the background photon flux $n_{\phi}^{(0)}$(Fig.1). This is satisfactory.
		Thus, this  novel property would provide an observable effect.\\
		
		\begin{figure}[!htbp]
			\centerline{\includegraphics[scale=0.5]{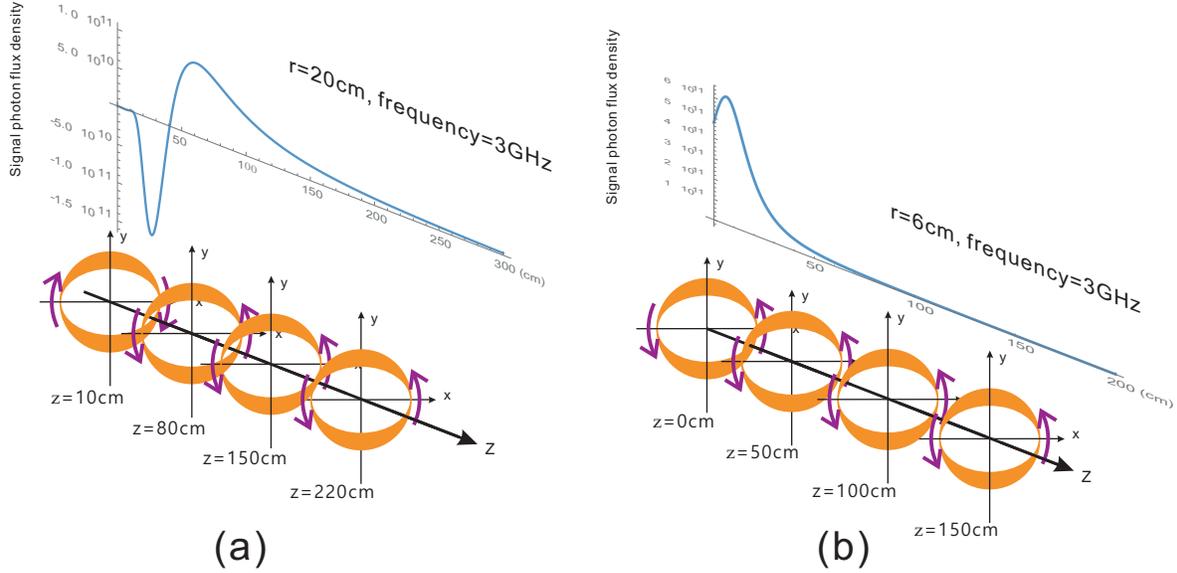}}\label{fig3}
			\caption{\footnotesize{Variation of rotational direction of quasi-B-mode along the propagation direction of HFGW. The
					$``z''$ means the distance to the minimal spot radius of the Gaussian beam, and ``r'' is the distance to the symmetrical axis of the
					Gaussian beam.  }}
		\end{figure}
		
		\indent Analytical expression  of the signal photon flux $n_{\phi-\otimes}^{(1)}$, Eq.(\ref{eq12}),  is a slow enough variational
		function in the propagating direction z of the HFGW. This means that ``rotation direction'' of $n_{\phi-\otimes}^{(1)}$ is
		as slow variational and it remains stable in the almost whole region of coherent resonance. For the HFGW of
		$\nu=3\times10^9Hz$(i.e., $\lambda_g=10cm$), $r=20cm$(distance to the symmetrical axis of Gaussian beam), the ``rotation direction'' of $n_{\phi-\otimes}^{(1)}$ keeps invariant
		in the first region of coherent resonance[the coherent resonance region keeping the right-handed rotation,
		i.e., from $z=10cm$ to $z=40cm$, see the curve in Fig.3(a)], and then, the ``rotation direction''
		will keep invariant
		in the next region of the coherent resonance[the coherent resonance region always keeping the left-handed rotation,
		i.e., in the region $z>40cm$[see the curve in Fig.3(a)]. For the case of $r=6cm$[see Fig.3(b)], the rotational direction has a better
		and more stable physical behavior, i.e., it will always keep left-handed rotational direction in the almost whole coherent
		resonance region.  In other words, the effective receiving area for the HFGW
		can be $\sim300~cm^2$. This means that it has enough large receiving surface to display the perturbative photon flux having $\nu_g=3GHz$.
		Especially, numerical
		calculation shows that this coherent effective resonance region will be enlarged as the frequency increases, so that the HFGWs having
		higher frequency will have a larger effective receiving surface(see Fig.4). Fig.4 shows relation between the frequencies of the
		HFGWs and the
		effective receiving surface.

		\begin{figure}[!htbp]
			\centerline{\includegraphics[scale=0.5]{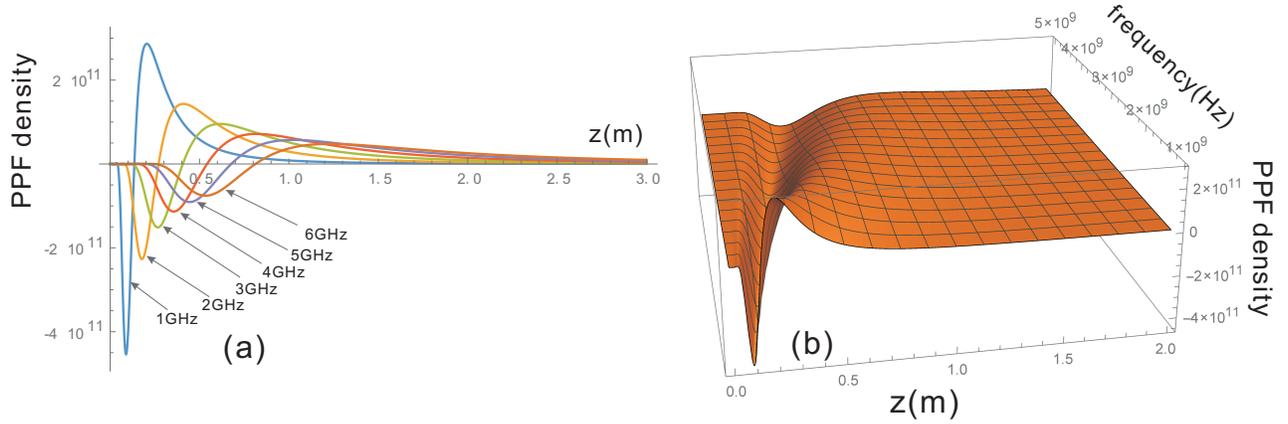}}\label{fig4}
			\caption{\footnotesize{(a)Perturbative photon flux(PPF) density for various frequencies. (b)Relationship between
					the perturbative photon flux(PPF) density with the frequency and ``z''(distance to the minimal spot radius of the Gaussian beam). The curve shows that the ``rotation direction'' of the perturbative photon flux(PPF)
					$n_{\phi-\otimes}^{(1)}$ in the EM synchro-resonance system would be more stable for the suitable region,
					and such distance can be effectively enhanced as the resonance frequency
					$\nu_g$ increases.}}
		\end{figure}

		Besides, because
		there are yet other different physical behaviours between the signal photon fluxes $n_{\phi-\otimes}^{(1)}$
		and the background photon flux, such as different propagating directions, distribution, decay rates(see,
		decay factors $exp(-\frac{2r^2}{W^2})$ of  $n_{\phi}^{(0)}$ in Eq.(\ref{eq7}) and $exp(-\frac{r^2}{W^2})$ of
		$n_{\phi-\otimes}^{(1)}$
		in Eqs.(\ref{eq12})), wave impedance(see below), etc in the special local regions, then it is always possible to
		distinguish the signal photon flux from the noise photons.\\
		
		\section{Displaying condition}
		\indent Since the signal photon fluxes are always accompanied by the noise photons, to identify the total signal photon flux at
		an effective receiving surface $\Delta S$, $n_{\phi(total)}^{(1)}\Delta t$ must larger than the total noise photon flux
		fluctuation at the receiving surface $\Delta s$. This displaying condition was discussed in Ref.\cite{26}, we shall not repeat it in detail here, and only give the main numerical calculation results. The displaying condition can be given by:\\
		\begin{eqnarray}
			\label{eq15}
			n_{\phi(total)}^{(1)}\Delta t~\geq~\sqrt{n_{\phi(total)}^{(0)}\Delta t},~
			then  ~\Delta t~\geq~n_{\phi(total)}^{(0)}/{[n_{\phi(total)}^{(1)}]}^2 = \Delta t_{min},
		\end{eqnarray}\mbox{}
		
		where $\Delta t_{min}$ is the requisite minimal accumulation time of the signal, and
		\begin{equation}
			\label{eq16}
			n_{\phi(total)}^{(1)}= \int\limits_{~~\Delta s} n_{\phi}^{(1)}ds,
			~n_{\phi(total)}^{(0)}= \int\limits_{~~\Delta s} n_{\phi}^{(0)}ds,
		\end{equation}\mbox{}

		are the total signal photon flux and the total noise photon flux passing
		through the receiving surface $\Delta s$, respectively.
		Actually,  there is a narrow frequency distribution of
		the Gaussian beam, and then the aimed signals caused by HFGWs also should not be monochromatic but with a sensitive
		frequency range. However, due to this frequency range is very short comparing to the HFGWs frequency band predicted by
		inflationary models or other scenarios, so we here calculate by a typical representative frequency instead of a frequency window.\\
		
		\begin{table}[!htbp]
			\footnotesize
			\caption{\label{tab:Table II}%
				Displaying condition for the HFGWs in some typical cosmological models and high-energy astrophysical process.}
			\begin{tabular}{ccccc}
				\hline
				\hline
				Amplitude(A)  &  $n_{\phi-\otimes(total)}^{(1)}(s^{-1}) $  &  $\Delta t_{min}(s)$  &Allowable upper limit
				&Possible verifiable cosmological    \\
				dimensionless &  &  & of noise photon flux$(s^{-1})$
				&models and astrophysical process \\
				\hline
				\\
				$10^{-23}$&$\sim1.6\times10^{9}$&$\sim10^4$&$2.8\times10^{22}$&Brane oscillation\cite{cqgF33},\\
				$10^{-27}$&$\sim1.6\times10^{5}$&$\sim10^6$&$2.8\times10^{15}$&Interaction of astrophysical plasma\\
				&&&& with intense EM radiation\cite{prd044017}\\
				\\
				$10^{-30}$&$\sim1.6\times10^{2}$&$\sim10^6$&$2.8\times10^{9}$&Pre-big-bang\cite{pr373,sciam290},\\
				&&&&Quintessential inflationary\cite{prd123511,cqg045004} or \\
				&&&&  upper limit of ordinary inflationarye\cite{0504018,084022} \\
				\hline
			\end{tabular}
		\end{table}
		\indent It should be pointed out that the background photon flux(in our synchro-resonance system, typical value of the Gaussian beam is ~10W) will be major source to the noise
		photon flux, i.e., other noise photon fluxes[e.g., shot noise, Johnson noise, quantization noise, thermal noise(if operation
		temperature $T<1k$), preamplifier noise, diffraction noise, etc.] are all much less than the background photon flux\cite{jmp498}. In other words, the
		Gaussian beam(the background photon flux) is likely to the dominant source of noise photons. Moreover, as mentioned earlier, the
		positions of maximum of the signal
		photon fluxes( $n_{\phi-\otimes}^{(1)}$ and  $n_{\phi-\otimes}^{(1)'}$, see Fig.2c and 2d) are just the zero value area of the
		background photon flux($n_{\phi}^{(0)}$, see Fig.1). Thus, major influence of the noise photon flux at such receiving surfaces would be from
		the background shot noise photon flux($\sim \sqrt{n_{\phi_{(total)}}^{(0)}}$) and not the background photon flux
		itself $n_{\phi_{(total)}}^{(0)}$. In this case,
		the relevant requirements to signal-to-noise ratio can be further relaxed.\\
		\indent Table 2 shows displaying condition of the HFGWs for some cosmological models and high-energy astrophysical process,
		where $n_{\phi-\otimes(total)}^{(1)}$ are the total signal photon fluxes at the receiving surface $\Delta s$ ($\phi=\pi/2$ or $3\pi/2$
		, $\Delta s\sim 3\times 10^{-2} m^2 $ ), which might be produced by the HFGWs in the Brane oscillation, quintessential inflationary, pre-big-bang models, and the interaction of high-energy plasma with EM radiation, and $n_{total}^{(0)}$ is allowable upper limit of the total noise photon flux at the surface
		$\Delta s$ for various values of the HFGW amplitudes and $\Delta t_{min}$, Eq.(\ref{eq15}), $\nu_e=\nu_g=3GHz$, the background
		static magnetic field $B^{(0)}$ is 10T, the interaction dimension $\Delta \textit{l}$ is 2m, the power of the Gaussian beam is $\sim10W$
		and operation temperature should be less than 1K.  Fortunately, one of institutes of our research team (High Magnetic Field Laboratory, Chinese Academic of Science) has been fully
		equipped with the ability to construct the superconducting magnet\cite{CASmagnet} (this High Magnetic Field Laboratory is also the superconducting magnet builder for
		the EAST tokamak for controlled nuclear fusion). The magnets can generate a static magnetic field with $\hat{B}^{(0)}=12~Tesla$ in an effective cross section of 80cm to 100cm at least,
		and operation temperature can be reduced to 1K even less. The superconducting static high field magnet will be used for our detection system. Then maximum of $n_{total}^{(0)}$  is $\sim10^{22}s^{-1}$ at the receiving
		surface $\Delta s$ of $\phi=\pi/4$, $3\pi/4$, $5\pi/4$ and $7\pi/4$, but it vanishes at $\phi=\pi/2$ and $3\pi/2$(see Fig.1).
		This means that at such surfaces even if the  noise photon flux reach up to the maximum ($\sqrt{n_{total}^{(0)}}\sim10^{11}s^{-1}$)
		of the background shot noise photon flux, then $\Delta t_{min}$ can be limited in $\sim10^6~seconds$ or less. For the HFGWs in the GHz band expected
		by the braneworld scenarios\cite{cqgF33},  both the maximum $\sqrt{n_{\phi}^{(0)}}|_{max}\sim10^{11}s^{-1}$ of the background shot
		noise photon flux or even the maximum($n_{\phi}^{(0)}|_{max}\sim10^{22}s^{-1}$) of the background photon flux itself are all less or much less
		than the allowable upper limit($\sim2.8\times10^{22}s^{-1}$, see Table 2) of noise photon flux. Thus, direct detection of the
		HFGWs\cite{cqgF33} in the braneworld scenarios would be quite possible due to larger amplitudes, higher frequencies, discrete
		spectral nature and extra polarization states for the K-K gravitons\cite{cqgF33,PhysRevD.88.064005,PRD104025}. Observation of
		the relic HFGWs predicted by the
		pre-big-bang\cite{pr373,sciam290}, the quintessential inflationary model\cite{prd123511,cqg045004} or the upper limit of the relic HFGWs expected by the
		ordinary inflationary models\cite{0504018,084022}, will face to enormous challenge, but it is not impossible.\\

		\begin{table}[!htbp]
			\footnotesize
			\caption{\label{Table3}%
				The perturbative photon fluxes (PPFs) $n_{\phi(total)}^{(1)}$ and the background photon fluxes (BPFs) $n_{\phi(total)}^{(0)}$ at the receiving $\Delta S$ where $\phi=89^\circ$ is azimuth angle in the cylindrical polar coordinates, and r is the distance to the symmetrical axis of the GB. We emphasize again that the important difference among the four main physical behaviors of $n_{\phi(total)}^{(1)}$ and $n_{\phi(total)}^{(0)}$: different angular distribution (see Fig.I and Fig.2(c),(d), the latter are just from the quasi-B-mode), different decay rate (see Eq.(7) and Eq.(12)),   very different wave impedance (see Tables IV and V) and different propagation directions in special regions. Importantly, in specific area of the detection system, the propagation direction of  $n_{\phi(total)}^{(1)}$ and  $n_{\phi(total)}^{(0)}$ are totally inverse each other, so in this case we can observe the signals using highly-oriented photon detector. Thus, only the signals will enter the photon detector and the background noise (of Gaussian Beam) can be effectively depressed. However, some other sources of noise should also be considered. In this table, by acceptable accumulation time of observation, we give the upper limit of these noise such as thermal noise, scattering and diffraction noise, Johnson noise, preamplifier noise and quantization noise\cite{jmp498}. It was shown\cite{jmp498} that such noise photon fluxes are less or much less than the upper limit of noise photon fluxes listed in this table.   }
			\begin{tabular}{ccccccc}
				\hline\hline
				HFGWE sources                      &          Position of the          & $n_{\phi(total)}^{(1)}$ & accumulation &  upper limit of noise   &  \\
				& receiving surface $\Delta S$ (cm) &       $(s^{-1})$        &   time (s)   & photon flux  $(s^{-1})$ &  \\ \hline
				Brane oscillation\cite{cqgF33}                     &          5cm$<$r$<$10cm           &    $2.836\times10^9$    & $\sim10^{4}$ &      $\sim10^{22}$      &  \\
				&          10cm$<$r$<$15cm          &   $3.235\times10^{8}$   & $\sim10^{4}$ &      $\sim10^{20}$      &  \\
				Quintessential inflationary \cite{prd123511,cqg045004} &                   &               & &           &  \\
				or   Pre-big-bang\cite{pr373,sciam290}         &      5cm$<$r$<$10cm                                &    283.6                         &       $\sim10^{4}$ &       $\sim10^{8}$     &  \\ \hline
				&                                   &
			\end{tabular}
		\end{table}

		\section{Wave impedance and wave impedance matching to the perturbative photon fluxes}
		\indent The wave impedance to an EM wave(photon flux) depends upon the ratio of the electric component to the magnetic component
		of the EM wave, and the wave impedance of free space to a planar EM wave is $377\Omega$\cite{Haslett}, and the wave impedance of copper
		to EM wave(photon flux) of $\nu=3\times10^9Hz$ is $0.02\Omega$\cite{Haslett}(see, Table 3). In fact, the wave impedance to the background photon
		flux(the Gaussian beam) and the planar EM waves in free space have the same order of magnitude($\sim377\Omega$). Unlike
		case of the wave impedance to the background photon flux, the ratio of the electric component of the perturbative photon flux(
		the signal photon flux) $n_{\phi-\otimes}^{(1)}$, Eq.(\ref{eq12}), to its magnetic component in selected wave zone of the
		synchro-resonance systems is much less than $377\Omega$.\\

		\begin{table}[!htbp]
			\footnotesize
			\caption{\label{tab:Table II}%
				The wave impedances to the EM waves(photon fluxes) in different materials\cite{Haslett}. This table shows that the wave impedance in
				the selected wave zone of the synchro-resonance system to the signal photon flux with $\nu_e=3\times10^9Hz$, is much less than
				$0.06\Omega$ of good conductors(e.g., copper), and even smaller than that of superconductor(see below).}
			\begin{tabular}{cccccc}
				\hline
				\hline
				Frequency& Wave impedance&Wave impedance&Wave impedance&Wave impedance&Wave impedance\\
				&Copper($\Omega$)&Silver($\Omega$)&Gold($\Omega$)&Superconductor($\Omega$)& Synchro-\\
				&&&&& resonance system($\Omega$)\\
				\hline
				\\
				$3\times10^9Hz$&0.060&0.063&0.046&$<10^{-3}$&$\sim10^{-4}$ or less\\
				\\
				\hline
			\end{tabular}
		\end{table}

		\begin{table}[!htbp]
			\small
			\caption{\label{tab:Table X}%
				The wave impedances in the selected wave zone of the synchro-resonance system to the transverse signal photon flux
				$n^{(1)}_{\phi-\otimes}$ and the background photon flux $n^{(0)}_{\phi}$. Here $\nu_e=\nu_g=3\times10^{12}Hz$, $A=10^{-23}$(e.g., the HFGWs
				in the braneworld model\cite{cqgF33})
				.}
			\begin{tabular}{cccc }
				\hline
				\hline
				Amplitude(A)& Position of & Wave impedance to & Wave impedance to  \\
				dimensionless&receiving surface(cm)& perturbative photon flux $n_{\phi-\otimes}^{(1)}(\Omega)$&   background photon flux $n_{\phi}^{(0)}(\Omega)$  \\
				\hline
				$10^{-23}$&$x=25,~y=5,~z=30$&$\sim3.04\times10^{-12}$&  \\
				$10^{-23}$&$x=30,~y=5,~z=30$&$\sim6.31\times10^{-9~}$&$\sim377$ \\
				$10^{-23}$&$x=35,~y=5,~z=30$&$\sim5.25\times10^{-5~}$&  \\
				$10^{-23}$&$x=25,~y=10,~z=30$&$\sim1.22\times10^{-11}$&  \\
				$10^{-23}$&$x=30,~y=10,~z=30$&$\sim2.53\times10^{-8~}$&$\sim377$ \\
				$10^{-23}$&$x=35,~y=10,~z=30$&$\sim2.11\times10^{-4~}$&  \\
				\hline
			\end{tabular}
		\end{table}
		
		\indent It is well known that energy of the electric components are far less than energy of the magnetic components for the EM
		waves(photon fluxes) propagating in good conductor and superconductor\cite{Haslett}. This means that the good conductor and superconductor
		have very low wave impedance, i.e., they have small Ohm losses for such photon fluxes. Then such EM waves(photon fluxes) are easy to
		propagate and pass through these materials. Fortunately, the signal photon flux in the typical wave zone of our synchro-resonance
		system has
		such property, i.e., the ratio of its electric component to the magnetic component is about 5 orders of magnitude less than that of
		background photon flux and other noise photons at least. This means that the signal photon flux $n_{\phi-\otimes}^{(1)}$ has
		very small wave impedance(see Table 4), and it would be easier to pass through the
		transmission way of the synchro-resonance system, i.e., the selected wave zone in the synchro-resonance system would be equivalent
		to a good ``superconductor'' to the perturbative photon flux.   Contrarily, the wave impedances to the background photon flux and other noise photons, are
		much greater than the wave impedance to the signal photon flux. I.e., Ohm losses produced by the background photon flux and
		the other noise photons would be much larger than Ohm losses generated by the signal photon flux in the photon flux receptors and
		transmission process. Therefore, the signal photon flux could be distinguished  from the background photon flux and other noise
		photons by the wave impedance matching(see Fig.5).\\
		\indent According to definition for the wave impedance\cite{Haslett} and Eqs.(\ref{eq2}),(\ref{eq6}),(\ref{eq11}) and (\ref{eq12}),
		we obtain the wave impedance Z of the typical receiving surface $\Delta s$ to the perturbative photon flux $n_{\phi-\otimes}^{(1)}$
		as follows:
		\begin{eqnarray}
			\label{eq17}
			\mathop{Z}=|\mu_0 { { \tilde {E}_y^{(1)}}}/{ \tilde {B}_z^{(0)}}|
			\approx\frac{\mu_0 A_{\otimes} { { \hat {B}^{(0)}}}\omega_g^2 W^2_0\Delta l}{4 \psi_0y}[1+(z/f)^{3/2}]\exp (\frac{r^2}{W^2}).
		\end{eqnarray}\mbox{}
		\begin{figure}[!htbp]
			\centerline{\includegraphics[scale=0.5]{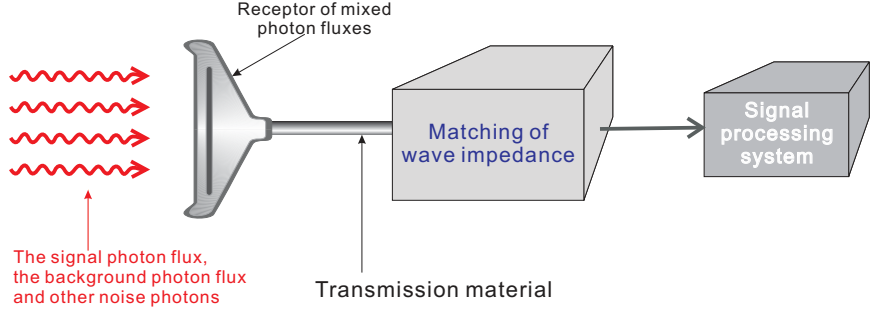}}\label{fig5}
			\caption{\footnotesize{Matching of wave impedance. This figure presents the basic scheme of the matching of wave impedance, to
					distinguish the signal photon flux from the background noise photon flux. The receptor can collect the mixed photon fluxes(the
					signal photon and the noise photon fluxes). However, the wave impedance($\sim10^{-4}\Omega$ or less, see Fig.6, Table 3 and 4)
					to the signal photon flux $n_{\phi-\otimes}^{(1)}$ is much less than that to the noise photon fluxes(including the background
					photon flux and other noise photons). The wave impedance matching and the signal processing systems can be only sensitive to
					the photon fluxes having the low wave impedance and not the photons with high wave impedance. Thus the signal photon flux would
					be selected and distinguished from the noise photons, due to their very different wave impedances. }}
		\end{figure}
		By using the typical parameters in the synchro-resonance system and in the typical cosmological models, i.e., $A_{\otimes}\sim
		10^{-23}$, $\nu_g=3THz$(e.g., the HFGWs in the braneworld mode\cite{cqgF33}),
		${\hat {B}^{(0)}}=10T$, $\psi_0=2.0\times10^3Vm^{-1}$(for the Gaussian beam of P=10W), $\Delta l=2m$, and selected wave zone:
		$y\in[5cm,10cm],~z\in[0,30cm],~x\in[5,30cm]$ for detection: some typical values of the wave impedance we obtained are listed in
		Table 4 and Fig.6. In the same way it can be shown that there are smaller wave impedance to the signal photon fluxes produced by the HFGWs
		expected by the pre-big-bang, and quintessential inflationary models(see Fig.6).
		\begin{figure}[!htbp]
			\centerline{\includegraphics[scale=0.85]{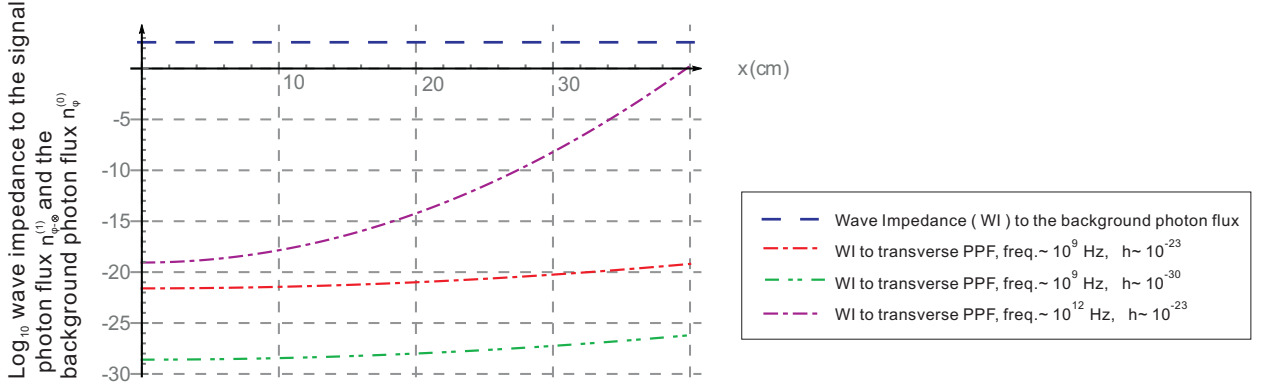}}\label{impedance}
			\caption{\footnotesize{It presents the comparison among wave impedances to the transverse perturbative photon fluxes
					$n_{\phi-\otimes}^{(1)}$ and to the background photon fluxes having different frequencies and amplitudes in the selected wave
					zone of the synchro-resonance system. Here x and y are distances to the longitudinal symmetrical surface(the yz-plane and
					xz-plane) of the Gaussian beam, respectively, and z is distance to the minimum spot radius of the Gaussian beam(see Fig.3), and $y=5cm$,
					$z=30cm$, $x\in[5,30cm]$. It is clear shown that the wave impedances($\sim10^{-4}\Omega$ or less) to the transverse
					perturbative photon flux $n_{\phi-\otimes}^{(1)}$ produced by the HFGW of $h=10^{-23}$, $\nu=3\times10^{12}Hz$(e.g., the HFGWs in
					the braneworld mode\cite{cqgF33}), are much less than that of the background photon flux $n^{(0)}_{\phi}$ in such region, and
					the wave impedances($\sim10^{-17}\Omega$ or less) to the perturbative photon flux $n_{\phi-\otimes}^{(1)}$ generated by the
					HFGWs of $h=10^{-30}$, $\nu=3\times10^9Hz$(e.g., the HFGWs in the pre-big-bang\cite{pr373,sciam290} or in the quintessential
					inflationary model\cite{prd123511,cqg045004}, are lot of orders of magnitude lower than that of the background photon flux
					in the wave zone.}}
		\end{figure}
		
		\begin{table}[!htbp]
			\footnotesize
			\caption{\label{Table6}%
				Similarity, complementarity and their difference between the two B-modes. }
			\begin{tabular}{ccc}
				\hline
				Properties&    B-mode in the CMB    &      Quasi-B-mode in the EM response       \\
				\hline
				Generation mechanism &Interaction of relic GWs with the CMB &EM resonance response to the HFGWs\\~\\
				Physical origin & Tensor perturbation&Tensor perturbation \\~\\
				Effect of available observation & B-mode polarization in the CMB&B-mode distribution of  perturbative \\
				& &    photon fluxes in the EM resonance\\~\\
				Intuitive image & Left-handed and right-handed swirls, & Left-handed and right-handed circular waves\\
				& \includegraphics[scale=0.4]{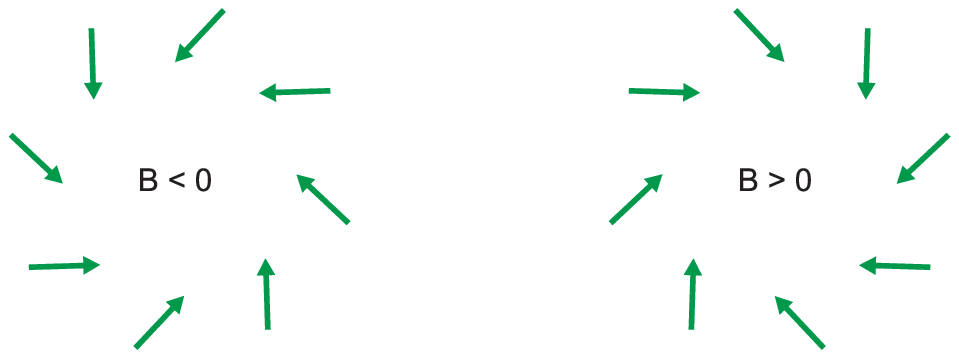} & \includegraphics[scale=0.4]{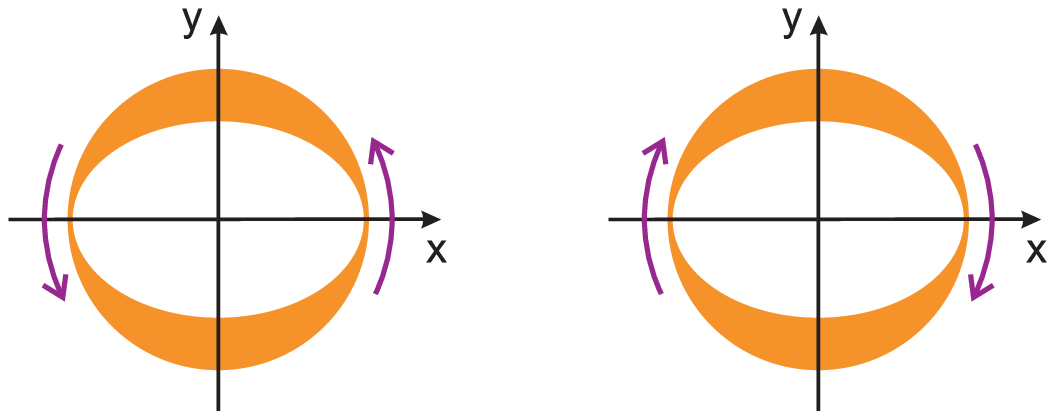} \\~\\
				Frequency bands& very-low frequency band &microwave frequency band \\
				& ($\sim10^{-16}~to~10^{-17}$Hz) &  ($\sim10^{9}~to~10^{12}$Hz)\\~\\
				Type of GWs &Primordial GWs in the&Primordial GWs in the\\
				& very low-frequency band  &high-frequency band and other HFGWs\\~\\
				Possible GW sources &Ordinary inflationary and & Quintessential inflationary,\\
				& other possible  inflationary & Pre-big-bang,brane oscillation and\\
				& &high-energy plasma vibration, etc. \\~\\
				Typical dimension of&Astrophysical scale & Typical laboratory dimension \\
				observation region& & \\~\\
				Major noise source&The cosmic dusts & The microwave noise photons \\
				& &inside the EM resonance system\\~\\
				Wave impedance to signals & $\sim377\Omega$ (the thermal photon   & $\sim10^{-4}\Omega$ or less (the perturbative  \\
				&distribution in the free space) & photon fluxes in the typical wave zone\\
				\hline
				
			\end{tabular}
		\end{table}

	\section{Concluding remarks}
	\indent(i)The B-mode in the CMB is from the interaction of the relic GWs with CMB, and this interaction produces the B-mode polarization
	in the CMB; the quasi-B-mode in the synchro-resonance system is from EM resonance response to the HFGWs;\\
	\indent(ii)The GW frequencies of the former are located in very low frequency band($\sim10^{-16} ~to ~10^{-17}Hz$), and the GW frequencies
	of the latter are occurred in typical microwave range($\sim10^{9}~to~10^{12}Hz$).\\
	\indent (iii)The B-mode of the former is distributed in astrophysical scale, and the quasi-B-mode of the latter is localized in typical
	laboratory dimension.\\
	\indent(iv)The major noise source in the former would be from the cosmic dusts, key noise in the latter is from
	the microwave photons inside the synchro-resonance system, which are almost independent of the cosmic dusts;\\
	\indent(v)Intuitive image of the former are the left-handed swirl and the right-handed swirl in the CMB(Fig.2b), and the physical
	picture of the latter are expressed as the ``left-hand circular wave'' and the ``right-hand circular wave'' distribution
	of the perturbative photon flux(Fig.2c and 2d);\\
	\indent (vi)The CMB displaying the B-mode are the EM waves(photon fluxes) in the free space, and in fact, it is also a
	thermal distribution of photons, and typical value of their wave impedance
	to
	the B-mode is $\sim377\Omega$\cite{Haslett}. Unlike the CMB, the wave impedance($\sim10^{-4}\Omega$ or less) to the signal photon flux
	in the
	typical wave zone of the synchro-resonance system is much less than that of the background photon flux and other noise
	photons. This means that the perturbative photon flux would be distinguished from the noise photons by the wave impedance
	matching. The similarity, complementarity and their difference between the two B-modes are listed in Table \ref{Table6}.\\
	\indent Notice, although the above two B-modes correspond to the different situations, their similarity and duality
	show that they are from the same physical origin: the tensor perturbation of the GWs and not the density perturbation, and only
	the GWs can generate such similarity and duality, and this is a very important difference to other perturbations and influences.\\
	\indent
	GWs in ordinary inflation model and the pre-big-bang model\cite{pr373,sciam290,nature990} involve issues of very early universe and
	the beginning of time;
	GWs in the braneworld model\cite{cqgF33, 12}  involves issues of
	the dimension of space, the multiverse, and direction of time arrow; GWs in the quintessential inflationary model\cite{prd123511, cqg045004}
	involve issues of the essence of dark energy, and GWs in high-energy astrophysical process\cite{prd044017}
	involve issues on the interaction mechanism of the interstellar plasma with intense EM radiation. These
	issues relate to important basic  questions: Does the universe have a beginning? If so, how did the universe
	originate? Was the big-bang the origin of the universe? Was our big-bang the only one? Does the multiverse exist?
	If so, can it be verified through scientific testing? Would quintessence be a serious candidate for dark energy?
	Could the interaction between astrophysical plasma and intense EM radiation provide stronger GW sources?\\
	\indent If the GWs are observed  in multiple frequency bands in the near future, and not only in the very
	low-frequency band($\nu\sim10^{-17}~to~10^{-16}Hz$), but also in low-frequency band($\nu\sim10^{-7}$ to $1Hz$),
	the intermediate-frequency band($\nu\sim1~to~10^{4}Hz$)
	and in the high-frequency band($\nu\sim10^{8}~to~10^{12}Hz$), and the observation results have highly self-consistence to
	the concrete cosmology parameters expected by certain cosmological model or a high-energy astrophysical scenario, then it will
	provide a stronger evidence for the model or the scenario. If not, the detection sensitivities or observation ways will need
	further improvement, or these models and scenarios will need to be corrected or will be ruled out. \\
	\indent Finally, it should be pointed out that, the HFGWs can also interact with galactic-extragalactic background magnetic fields, and then lead to EM signals with the same frequency as the HFGWs. Although the galactic-extragalactic background magnetic fields are very weak $\sim10^{-9}$ to  $\sim10^{-11}$ T, the huge propagation distance could result in a useful spatial accumulation effect in the propagational direction\cite{PRD104025}, due to the same propagation velocities of HFGWs and EM signals. This may lead to a possibly observable effect on the Earth. Fortunately, such EM signals ($10^8$ to $10^9$Hz) sit in the detection frequency band of FAST (Five-hundred-meter Aperture Spherical Telescope) which is expected  to be completely constructed in 2016 in Guizhou province, China. Therefore, the observation by FAST, detection of the HFGWs by our resonance detection system, and our cooperation with FAST can be strongly complementary. Those consequent works will be carried out in the near future.

\begin{acknowledgments}
This project is supported by the National Natural Science Foundation of China (No.11375279, No.11605015 and No.11205254), the Foundation of China Academy of Engineering Physics (No.2008 T0401 and T0402), the Fundamental Research Funds for the Central Universities(No.106112016CDJXY300002 and 106112015CDJRC131216), and the Open Project Program of State Key Laboratory of Theoretical Physics Institute of Theoretical Physics, Chinese Academy of Sciences, China (No.Y5KF181CJ1).
\end{acknowledgments}

\bibliographystyle{apsrev4-1}
\bibliography{3DSRReferenceData}
\end{document}